# Perfect Coulomb drag in a dipolar excitonic insulator

Phuong X. Nguyen[1], Liguo Ma[1], Raghav Chaturvedi[1], Kenji Watanabe[2], Takashi Taniguchi[2], Jie Shan[1,3,4]*, Kin Fai Mak[1,3,4]*

[1]School of Applied and Engineering Physics, Cornell University, Ithaca, NY, USA
[2]National Institute for Materials Science, Tsukuba, Japan
[3]Laboratory of Atomic and Solid State Physics, Cornell University, Ithaca, NY, USA
[4]Kavli Institute at Cornell for Nanoscale Science, Ithaca, NY, USA

*Email: jie.shan@cornell.edu; kinfai.mak@cornell.edu

These authors contributed equally: Phuong X. Nguyen, Liguo Ma, Raghav Chaturvedi

**Excitonic insulators (EIs), arising in semiconductors when the electron-hole binding energy exceeds the band gap[1–3], are a solid-state prototype for bosonic phases of matter[4–12]. Unlike the charged excitations that are frozen and unable to transport current, the neutral electron-hole pairs (excitons) are free to move in EIs. However, it is intrinsically difficult to demonstrate exciton transport in bulk EI candidates[13–18]. The recently emerged dipolar EIs based on Coulomb-coupled atomic double layers[4,19–24] open the possibility to realize exciton transport across the insulator because separate electrical contacts can be made to the electron and hole layers. Here we show that the strong interlayer excitonic correlation at equal electron and hole densities in the MoSe$_2$/WSe$_2$ double layers separated by a 2-nm barrier gives rise to 'perfect' Coulomb drag. A charge current in one layer induces an equal but opposite drag current in the other. The drag current ratio remains above 0.9 up to about 20 K for low exciton densities. As exciton density increases above the Mott density[25], the excitons dissociate into the electron-hole plasma abruptly, and only weak Fermi liquid frictional drag is observed. Our experiment moves a step closer to realizing exciton circuitry[26] and superfluidity.**

**Main**

The dipolar excitonic insulator (EI) of interest is realized in double layer structures made of two semiconducting transition metal dichalcogenide (TMD) monolayers that support a type-II band alignment[21] (Fig. 1a). The two TMD monolayers, which are separated by a thin dielectric spacer, are separately contacted by metallic electrodes and form the two plates of a parallel-plate capacitor (Fig. 1b). The application of an interlayer bias voltage, $V_b$, splits the electron and hole chemical potentials by $eV_b$, and reduces the charge gap to $\mathcal{E}_G - eV_b$ for quasiparticle excitations from the zero-bias band gap, $\mathcal{E}_G$, of the double layer (Fig. 1a). (Here $e$ is the elementary charge.) When the charge gap becomes smaller than the exciton binding energy, $\mathcal{E}_G - eV_b < \mathcal{E}_b$, spontaneous formation of dipolar (interlayer) excitons is favorable. An EI has been demonstrated by recent thermodynamic measurements[21,24]. The separate electron- and hole-contacts to the EI also provide an electrical reservoir for dipolar excitons with continuously tunable density, $n_X$, and chemical potential, $eV_b$ (Ref. [19,20]).

The Coulomb drag effect is a sensitive probe of the correlation effect between two spatially separated conductors[27]. The Coulomb interactions between the two layers here enable an electric current flowing in one layer ($I_{drive}$) to induce a voltage drop ($V_{drag}$) in the other (Fig. 1d). If the second layer is part of a closed circuit, a net current ($I_{drag}$) will flow in that circuit (Fig. 1c). The drag current is typically much smaller than the drive current because the Coulomb interactions are heavily screened. Here we show essentially perfect Coulomb drag up to about 20 K in the dipolar EI based on TMD electron-hole double layers. Such a large drag current ratio ($I_{drag}/I_{drive}$) has been observed only in the context of exciton transport in the quantum Hall regime demonstrated in Coulomb-coupled quantum wells[28–30] and graphene double layers[31–34]. In contrast to the high magnetic fields required in these earlier studies, the strong exciton binding under zero magnetic field and the dispersive excitonic band in TMD double layers present a new regime for the studies of exciton transport[35–37].

**Device operation and excitonic insulator**
Our sample consists of Coulomb-coupled $MoSe_2$/$WSe_2$ double layers with nearly symmetric top and bottom gates (Fig. 1b). Details on the device design and fabrication are described in Methods. These devices are divided into the channel and contact regions, in which the two TMD monolayers are separated by thin (5-6 layers) and thick (10-20 nm) hexagonal boron nitride (hBN) spacers, respectively. They are angle-misaligned to further suppress interlayer tunneling in the channel region. The $MoSe_2$ (Mo) and $WSe_2$ (W) layers are contacted by bismuth and platinum electrodes to achieve electron and hole transport measurements, respectively. The interlayer bias voltage $V_b$ is applied to inject electron-hole pairs into the channel. The top and bottom gate voltages ($V_{tg}$ and $V_{bg}$, respectively) further allow independent control of the electron-hole density imbalance in the double layer through the symmetric combination, $V_g = (V_{tg} + V_{bg})/2$, and the perpendicular electric field through the antisymmetric combination, $\Delta = (V_{bg} - V_{tg})/2$. In the channel region, the electric field reduces the band gap; but in the contact regions, because of the much thicker hBN spacer, $\Delta$ is able to heavily electron- and hole-dope the Mo and W layers, respectively, to substantially reduce the metal-semiconductor junction resistances. Junction resistances on the order of several kΩ's at 1.5 K can be achieved. The contact regions are thus charge reservoirs for efficient exciton injection into the channel region[21].

We identify the EI phase of the double layer using the capacitance measurements as reported in an earlier study[21] (Methods, Extended Data Fig. 2). Both the penetration ($C_P$) and interlayer ($C_I$) capacitances are measured as a function of $V_g$ and $V_b$ at $\Delta = 5.5$ V and temperature $T \approx 1.5$ K. They characterize the charge and exciton compressibilities[21], respectively. Figure 2a illustrates $C_P$ normalized by its geometric value ($C_{gg}$). The large penetration capacitance at small bias voltages marks the *ii* region with both layers being charge-neutral. Its boundary, denoted by two black dashed lines, traces the band edge of the electron and hole layers. Particularly, the band edge of the hole layer is fixed at $V_g = 0$ because the layer is grounded. Varying the gate voltage across each boundary introduces doping in one of the layers (the *pi* and *in* regions). At sufficiently large bias voltages, both layers become doped (the *pn* region). Further, the simultaneous interlayer capacitance measurement (Extended Data Fig. 2b) identifies the dash-dotted line, above which the electron-hole pairs are injected into the channel. The enclosed triangular region is therefore an EI, which at low temperatures is charge insulating but hosts an equilibrium dipolar exciton fluid. The exciton binding energy $\varepsilon_b \approx 30$ meV is approximately given by the difference between the bias voltage at the tip of the triangle (0.52 V), beyond which the excitons dissociate into free electrons and holes, and the base of the triangle (0.49 V), at which excitons are just injected into the double layer channel. The channel band gap $\varepsilon_G$ is reduced from the intrinsic value of 1.6 eV to 0.52 eV by antisymmetric gating $\Delta$.

**Perfect Coulomb drag**

We study Coulomb drag in our device using the drag counterflow geometry illustrated in Fig. 1c. Specifically, we drive an electron current in the Mo layer using a small ac in-plane bias (5-10 mV) and measure the drag current in the W layer (Methods). Figure 2b and 2c show, respectively, the $V_g$- and $V_b$-dependence of the drive and drag currents at 1.5 K (see Extended Data Fig. 3 for results from driving the W layer). As expected, a finite $I_{\text{drive}}$ is observed whenever the Mo layer is turned on, that is, in the *in*, *pn* and EI regions; and the current increases with increasing electron doping. The discrepancy between the transport and capacitance results near the Mo band edge in the *in* region is likely related to the large contact resistance in the Mo layer in the device (see Methods and Extended Data Fig. 5). In contrast, a $I_{\text{drag}}$ hot spot is observed only near the EI region; it decays rapidly with both increasing $V_b$ and $V_g$. The former increases the exciton density, and the latter, the electron-hole density imbalance. In addition, $I_{\text{drag}}$ always flows in the opposite direction of $I_{\text{drive}}$ (Extended Data Fig. 4).

Below we focus on the case of equal electron and hole densities in the double layer (along the white dashed line in Fig. 2a). The more general case of finite electron-hole density imbalance involving a Bose-Fermi mixture[38,39] and screened excitonic interactions will be examined in future studies. Figure 3a shows the current amplitude in two layers at 1.5 K as a function of $V_b$ or electron-hole pair density, $n_P$. The pair density is obtained from the interlayer capacitance measurement, $n_P = \int \frac{C_I}{e} dV_b$ (Extended Data Fig. 2c). The drive current turns on sharply at $V_b \approx (\varepsilon_G - \varepsilon_b)/e$ when excitons are first injected into the channel, followed by a moderate increase with increasing $V_b$. The small deviations of the drive current from a smooth monotonic $V_b$-dependence are device specific; they are likely related to the junction imperfections and require systematic studies to understand the origin. Remarkably, $I_{\text{drag}}$ is nearly identical to $I_{\text{drive}}$ (but with opposite sign) until the bias reaches a threshold, $V_b \approx \varepsilon_G/e$, beyond which $I_{\text{drag}}$ vanishes abruptly. The corresponding pair density is $n_M \approx 3.2 \times 10^{11}$ cm$^{-2}$. Figure 3a also shows that the tunneling current $I_{\text{tunnel}}$ between the two layers is negligible and cannot compromise the drag measurements.

The lower panel of Fig. 3a shows the measured drag current ratio at 1.5 K. It displays 'perfect' Coulomb drag in the EI region and negligible Coulomb drag in the free electron and hole region with $n_P > n_M$. The temperature and pair density dependence of the drag current ratio is summarized in Fig. 3b, and representative density line cuts are illustrated in Fig. 3c. In general, Coulomb drag decreases with increasing temperature and pair density with the exception of the region near $n_M$ at low temperatures, where a non-monotonic temperature dependence is observed. Essentially perfect Coulomb drag ($I_{\text{drag}}/I_{\text{drive}} > 0.9$) persists up to about 20 K at low pair densities.

When exciton transport dominates over the free carrier transport in the double layer channel, perfect Coulomb drag is expected because a steady current of electrons driven through one layer has to be accompanied by an equal current of holes in the other layer[28]. In contrast, when the electrons and holes are unbound, only small frictional drag is expected[27]. In the case of Fermi liquids, the phase space of electrons and holes available for Coulomb drag increases with temperature, and the drag current ratio scales quadratically with temperature at low temperatures[27]. The observed 'perfect' Coulomb drag in our experiment therefore demonstrates the dominance of exciton transport in the EI region at low temperatures. The temperature dependent Coulomb drag in this regime can be captured by the exciton thermal dissociation effect[40]. In this simple model (solid lines in Fig. 3c), the frictional drag and the contact resistance are negligible, and the exciton to charge carrier mobility ratio is the only fitting parameter and assumed to be temperature independent (Methods). The excitons can also be ionized into an electron-hole plasma at low temperatures by the state filling effect when the pair density exceeds the Mott density[40]. The latter corresponds to the abrupt vanishing of the drag current near $n_M$ at low temperatures, and $n_M$ is the Mott density. In the high-density electron-hole plasma regime, Coulomb drag increases with temperature at low temperatures by the phase space argument ($n_p = 5 \times 10^{11}$, Fig. 3c).

**Excitonic insulator-to-metal transition**

We measure the drag resistance of our device to directly probe the transition from an EI to an electron-hole plasma using the open-circuit geometry shown in Fig. 1d. Here we bias $I_{\text{drive}}$ in the Mo layer and measure the drag voltage drop in the W layer, $V_{\text{drag}}$, by a voltmeter with 100 MΩ input impedance. In this measurement geometry, the drag resistance, $R_{\text{drag}} = V_{\text{drag}}/I_{\text{drive}}$, reflects the charge resistance of the double layer (see the circuit model in Extended Data Fig. 1d and Methods). It is ultimately connected to the drag current measurement above. In the EI region at low temperatures (where the exciton conductivity far exceeds the charge conductivity and the frictional drag conductivity), the two quantities are simply related, $\frac{I_{\text{drag}}}{I_{\text{drive}}} = \frac{1}{1+R_{\text{WC}}/R_{\text{drag}}}$, via the contact resistance ($R_{\text{WC}}$) in the drag layer (see Extended Data Fig. 1f). The good agreement between the measured drag current ratio and the inferred one from $R_{\text{drag}}$ using $R_{\text{WC}} \approx 7$ kΩ in Extended Data Fig. 7 supports the validity of the relation for a wide range of temperatures and pair densities. The contact resistance value is independently verified by a two-terminal resistance measurement in the W layer (Extended Data Fig. 5).

Figure 4a shows the temperature dependence of $R_{\text{drag}}$ at varying pair densities. As temperature decreases, $R_{\text{drag}}$ increases for small $n_\text{P}$'s (an insulating behavior) and decreases for large $n_\text{P}$'s (a metallic behavior). A variation in $R_{\text{drag}}$ by nearly six orders of magnitude is observed at 1.5 K. On the metallic side, a Fermi liquid frictional drag, $R_{\text{drag}} \approx AT^2$, is observed in the low-temperature limit [27,41] (Extended Data Fig. 6). And a non-monotonic temperature dependence is also visible near the Mott density $n_\text{M}$.

We analyze the observed insulator-to-metal transition by identifying the resistance at the critical density, $R_\text{c}$, as the one that displays a power-law temperature dependence[42] (dashed line). The corresponding density agrees well with the Mott density $n_M$. We then normalize the temperature dependent $R_{\text{drag}}$ at other densities by $R_\text{c}$. We are able to collapse all the temperature dependent resistances $R_{\text{drag}}/R_\text{c}$ into two groups, exhibiting an insulating and metallic behavior, respectively, by normalizing $T$ using a density-dependent temperature scale, $T_0$ (inset, Fig. 4a). The temperature scale $T_0$ vanishes continuously towards the Mott density $n_M$ from both sides (Extended Data Fig. 9). On the insulating side, it is well correlated with the EI charge gap ($\approx \int (C_\text{P}/C_\text{gg}) \, dV_\text{g}$) obtained from the penetration capacitance measurement. The charge gap summarized in Fig. 4b carries a relatively large systematic uncertainty induced by the finite ac bias voltage in the penetration capacitance measurement. On the metallic side, we summarize in Fig. 4b the coefficient $A$ from fitting the temperature dependent $R_{\text{drag}}$ to $AT^2$ at low temperatures. The coefficient increases by three orders of magnitude when $n_\text{P}$ approaches $n_\text{M}$ from above. The combined results of a continuously vanishing charge gap (and $T_0$) and a diverging $A$ coefficient when the pair density approaches the Mott density from two sides suggest a possible continuous excitonic insulator-to-metal transition[43].

We make a further interesting observation that the coefficient $A$ from our experiment scales with pair density as $A \propto (n_\text{P} - n_\text{M})^{-3}$ (solid line, Fig. 4b). This resembles, but at the same time differs from, the theoretical prediction of $A \propto (n_\text{P})^{-3}$ for frictional drag from the interlayer Coulomb scattering[27,44]. The reset of the pair density by the Mott density suggests a vanishing effective electron/hole Fermi energy (or a diverging electron/hole effective mass) at $n_\text{M}$ (Ref. [27,44]). The reset is beyond the current understanding. Future theoretical studies taking into account the BEC-BCS crossover[45] and the possible Mott transition at $n_\text{M}$ are required to understand the density dependence of $A$ as well as the non-monotonic temperature dependence of $R_{\text{drag}}$ near $n_\text{M}$ (BEC and BCS stand for Bose-Einstein condensation and Bardeen-Cooper-Schrieffer, respectively).

## Conclusions

We have demonstrated essentially perfect Coulomb drag up to about 20 K in a gate-tunable excitonic insulator realized in TMD atomic double layers separated by a thin hBN barrier. The drag resistance measurement has further revealed an excitonic insulator-to-metal transition at the Mott density. These studies are enabled by successfully establishing separate electrical contacts to the electron and hole components of the excitonic insulator. The perfect Coulomb drag here demonstrates pure exciton transport across the excitonic insulator. The relatively large contact resistance of the present device limits the ability to settle whether the exciton transport here is dissipationless, that is, whether there is exciton superfluity. Future multiple exciton terminal measurements that can drive an exciton current and simultaneously measure the exciton chemical potential drop along the channel should be able to set strict limits on any dissipation occurring in the exciton channel.

## Methods
### Device design and fabrication

In Extended Data Fig. 1a,b, we show an optical image and a schematic for our dual-gated $WSe_2$/hBN/$MoSe_2$ transport device to illustrate the basic design geometry. The top and bottom gates consist of few layer graphite gate electrodes and ~10 nm thick hBN dielectrics. The overlap region between the top gate, bottom gate and the TMD double layer defines the measurement channel and the two exciton contacts, which are shaded red and blue, respectively, in the optical image. In the channel region (region I in the schematic), the hBN barrier is around 1.5-2 nm thick (measured by atomic force microscopy); this thickness is optimized to suppress the interlayer tunneling current while maintaining a large exciton binding energy (~30 meV). In the exciton contact regions (region II in schematic), the two thick hBN barriers (10-20 nm) result in heavily doped regions of free electrons and holes (no exciton binding) under a perpendicular electric field; these regions serve as electrical reservoirs for efficient exciton injection and absorption in and out of the channel.

Outside the channel region, the flow of exciton must be converted into flows of free electron and holes in the electrical wires. It is therefore crucial to have good electrical contacts to the monolayer TMDs for simultaneous electron and hole transport at low temperatures (down to 1.5 K). We chose platinum (Pt) that has a large work function to match the valence band edge of monolayer $WSe_2$ (Ref. [46]); this consistently yields < 10 kΩ contact resistance at temperatures down to 1.5 K. For the monolayer $MoSe_2$ conduction band edge, we used bismuth (Bi) [47], a semimetal with a small work function; it can also achieve < 10 kΩ contact resistance at T = 1.5 K but has a lower yield compared to the Pt contacts.

In addition, we designed a sample geometry such that the metal-TMD contact regions (region III in the schematic) are gated by only one gate (as seen in the optical image, the $WSe_2$ contact pins W10 to W14 are covered only by the top gate, and the $MoSe_2$ contact pins Mo6 to Mo9 are covered only by the bottom gate). With this specific design and the large perpendicular electric field from the antisymmetric gate voltage Δ = 5.5 V (large negative top gate and positive bottom gate voltages), we guarantee that the metal-TMD contact regions are heavily doped (more than the exciton contact regions) by holes for Pt-WSe$_2$ and by electrons for Bi-MoSe$_2$ for low contact resistances. Note that it remains challenging to obtain a device with multiple good contacts in each layer and a good geometry for exciton injection, mainly due to the variability of the mechanical stacking and microfabrication process detailed below. For our main device at T = 1.5 K, all five WSe$_2$ contacts (W10 to W14) are around 5-7 kΩ in contact resistance; the two MoSe$_2$ contacts (Mo8 and Mo9) are around 5 kΩ; the MoSe$_2$ contact Mo7 is around 3 MΩ; and the MoSe$_2$ contact Mo6 was mechanically broken during fabrication.

We fabricated our device using the layer-by-layer dry transfer technique reported by Ref. [48]. The flakes were obtained from mechanical exfoliation of bulk TMD crystals (HQ Graphene) onto Si/SiO$_2$ substrates. The layers were then picked up sequentially at ~50 °C using a polymer stamp made of a thin layer of polycarbonate on a polypropylene-carbonate-coated polydimethylsiloxane block. At each step, careful alignment of the layers (up to 1 um lateral precision) is essential to achieve the ideal device geometry, especially for the exciton contacts. The completed stack (9 layers in total) was then released onto an amorphous quartz substrate with pre-patterned Pt electrodes to make contacts to the WSe$_2$ monolayer and to the gate electrodes. The residual polycarbonate film was removed in chloroform and isopropanol. Amorphous quartz substrates were chosen for minimal parasitic capacitance background. The top graphite gate was then etched to define multiple contacts for the WSe$_2$ layer. We used standard e-beam lithography followed by reactive-ion etching by oxygen plasma (Oxford Plasmalab80Plus). Lastly, we made Bi contacts to the MoSe$_2$ layer by a second e-beam lithographic step to define the contact regions, followed by Bi deposition in a thermal evaporator at a rate of 0.5 Å/s. Since Bi oxidizes in air, we evaporated the metal as the last step to guarantee the best MoSe$_2$-Bi interface; we also cannot use the standard pre-patterned electrodes scheme for WSe$_2$ on Pt.

**Coulomb drag measurements**
We examine the exciton transport properties using two different configurations: the open-circuit and closed-circuit Coulomb drag measurements. Extended Data Fig. 1c,d show the circuit diagrams. In order to choose the driving layer, we first performed resistance measurements on each TMD layer separately to determine the contact resistances for each layer (Extended Data Fig. 5). We chose the Mo layer as the driving layer because of its much larger contact resistance (up to 3 MΩ) compared to that in the W layer (about 7 kΩ). As a result of the large Mo contact resistance, a proper closed-circuit Coulomb drag measurement cannot be performed if we chose the W layer as the driving layer because it would act like an open-circuit element.

We applied a 10 mV (RMS) AC bias voltage at 7.33 Hz to drive the Mo layer through a 1:1 voltage transformer, which was connected to a 10 kΩ potentiometer to distribute the AC voltage on the two ends of the driving layer. This can reduce the AC coupling between the TMD layers by minimizing the AC interlayer potential difference in the channel. A DC voltage source was also connected to middle of the potentiometer to apply the interlayer bias voltage $V_b$ between the driving (Mo) and the drag (W) layers. We further connected a 150 kΩ resistor in series with the driving layer; we measured the voltage drop across this resistor to obtain the driving current.

In the closed-circuit measurement, we shorted the drag layer so that a drag current can flow. We used a current preamplifier and a SR830 lock-in amplifier to measure the drag current. In the open-circuit measurement, we connected the two ends of the drag layer to a voltage amplifier with 100 MΩ input impedance and directly measured the voltage drop between the two ends using a SR830 lock-in amplifier in order to obtain the drag resistance.

We carried all of the measurements in a closed-cycle He4 cryostat (Oxford TeslatronPT) down to 1.5 K temperature. The measurement results are largely independent of the excitation amplitude (3–20 mV) and frequency (7 Hz to 37 Hz).

**Capacitance measurements**
Details of the capacitance measurements have been reported by Ref. [21]. In short, we measured the penetration capacitance of the bilayer system to determine the electrostatic phase diagram. We applied a 5 mV (RMS) AC bias voltage at 737 Hz to the bottom gate, which induces a change in the carrier density, $dn$, in the top gate. This density change was measured by a low-temperature capacitance bridge based on a GaAs high-electron-mobility-transistor (HEMT). We also measured the interlayer capacitance between the TMD layers to determine the pair density in the channel. We applied a 5 mV (RMS) bias voltage at 1.577 kHz to the Mo layer and collected the displacement current from the W layer. The in-phase component is the tunneling current while the 90-degree out-of-phase component is proportional to the change in the pair density, $dn_P$; the complex current was measured by the low-temperature HEMT transimpedance amplifier and the SR830 lock-in amplifier.

**Modeling the Coulomb drag ratio and drag resistance**
Extended Data Fig. 1e shows a circuit model that describes the electron, hole and exciton current flows in the bilayer. For a given applied in-plane electric field $E_e$ in the Mo layer, the current density $J_e$ in the same layer has two contributions, one due to free electrons with conductivity $\sigma_e = n_e e \mu_e$ and the other due to excitons with conductivity $\sigma_X = n_X e \mu_X$. Similarly, the current density in the drag layer $J_h$ has contributions from the free holes ($\sigma_h = n_h e \mu_h$) and the excitons ($\sigma_X$). [Here $n_{e(h)}$ and $\mu_{e(h)}$ denote the free electron (hole) density and mobility, respectively.] The transport in the bilayer system can be described by a conductivity tensor $\Sigma$ (Ref. [49]) as:

$$\begin{pmatrix} J_e \\ J_h \end{pmatrix} = \Sigma \begin{pmatrix} E_e \\ E_h \end{pmatrix} = \begin{pmatrix} \sigma_X + \sigma_e & -\sigma_X - \sigma_D \\ -\sigma_X - \sigma_D & \sigma_X + \sigma_h \end{pmatrix} \begin{pmatrix} E_e \\ E_h \end{pmatrix}. \tag{1}$$

The drag conductivity $\sigma_D$ describes the frictional drag from the free electrons and free holes. For the open-circuit measurement, we have $J_h = 0$ and can obtain the drag resistance (for a channel length approximately equal to the channel width, see Extended Data Fig. 1a) as:

$$R_{\mathrm{drag}} \approx \frac{E_h}{J_e} = \frac{\sigma_X + \sigma_D}{\det \Sigma}. \tag{2}$$

For the closed-circuit measurement, the drag layer acts like a battery with an external load resistance equal to the total contact resistance in the W layer, $R_{WC}$; we have $E_h \approx -J_h R_{WC}$ for a channel length approximately equal to the channel width. We can obtain an expression for the drag current ratio:

$$\left|\frac{J_h}{J_e}\right| \approx \frac{\sigma_X + \sigma_D}{\sigma_X + \sigma_e + (\det \Sigma) R_{WC}} = \frac{1}{\frac{\sigma_X + \sigma_e}{\sigma_X + \sigma_D} + \frac{R_{WC}}{R_{drag}}}. \tag{3}$$

Note that the final result on the right of Eqn. (3) is exact and independent of the channel geometry because the geometry dependence is cancelled in the factor $\frac{R_{WC}}{R_{drag}}$. Finally, by assuming $\sigma_X \gg \sigma_e, \sigma_D$, which is a good approximation for $n_P < n_M$ and at low temperatures, we get $\left|\frac{J_h}{J_e}\right| \approx \frac{1}{1 + \frac{R_{WC}}{R_{drag}}}$ as discussed in the main text; this approximate result is equivalent to a lumped circuit diagram shown in Extended Data Fig. 1f. Based on the measured drag resistance and the drag current ratio, we can confirm that this approximate expression works well for a wide range of exciton densities and temperatures for $R_{WC} \approx 7\ k\Omega$ (Extended Data Fig. 7). Moreover, independent two-terminal measurements verify $R_{WC} \approx 7\ k\Omega$ (Extended Data Fig. 5).

**Temperature dependence of the drag current ratio**
In the EI phase ($n_P < n_M$), we can assume small frictional drag ($\sigma_X \gg \sigma_D$) and small contact resistance ($R_{WC} \ll \sigma_e^{-1}$) and obtain

$$\left|\frac{J_h}{J_e}\right| \approx \frac{\sigma_X}{\sigma_X + \sigma_e} = \frac{n_X}{n_X + n_e(\mu_e/\mu_X)}. \tag{4}$$

We can use the Saha equation to estimate the free carrier density due to thermal dissociations of excitons at finite temperatures [40,49]:

$$\frac{n_e^2}{n_X} = \frac{k_B T}{2\pi\hbar^2/m^*} e^{-\varepsilon_b/k_B T}, \tag{5}$$

where $m^*$ is the exciton effective mass. For a fixed pair density $n_P = n_e + n_X$, the exciton binding energy $\varepsilon_b$ can be obtained from penetration capacitance measurements (Extended Data Fig. 8). We can then solve for $n_e$ and $n_X$ using Eqn. (5) and fit the experimental temperature dependence of the drag current ratio using Eqn. (4). The only fitting parameter is the mobility ratio ($\mu_e/\mu_X$), which is 2, 2.8 and 3.3 for $n_P = 1, 2$ and $3 \times 10^{11}$ cm$^{-2}$, respectively.

**Acknowledgements**
We thank Allan MacDonald, Liang Fu, Sankar Das Sarma, Peter Littlewood and Erich Mueller for many helpful discussions.

**Figures**

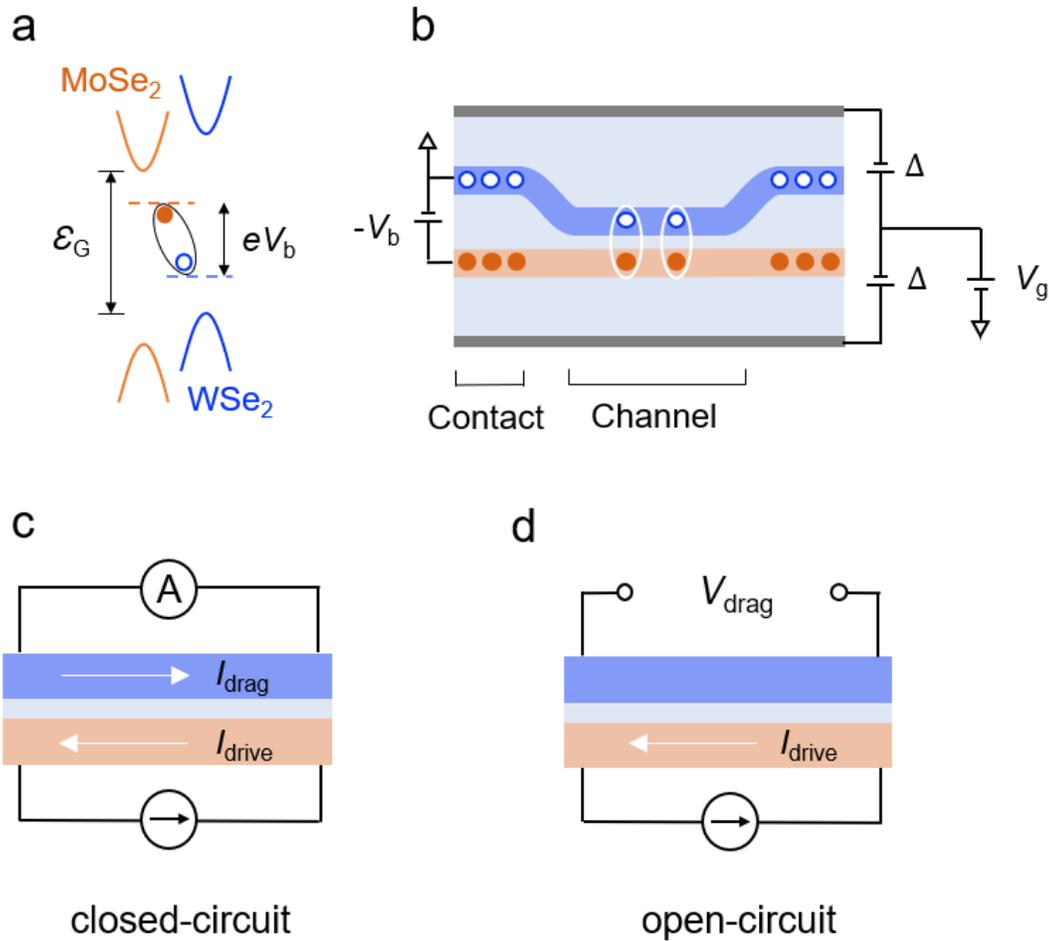

**Figure 1 | Coulomb drag measurements in a dipolar excitonic insulator. a,** Type-II band alignment of the MoSe$_2$/WSe$_2$ heterostructure with an interlayer band gap $\mathcal{E}_G$. An interlayer bias voltage $V_b$ is applied to separate the electron (solid dot) and hole (empty dot) chemical potentials and to reduce the charge gap of the bilayer to $\mathcal{E}_G - eV_b$. An EI spontaneously forms when $V_b > (\mathcal{E}_G - \mathcal{E}_b)/e$. **b,** Cross-section schematic of the dual-gated WSe$_2$/hBN/MoSe$_2$ transport device. The W layer, Mo layer, graphite gates, and hBN dielectrics are colored in blue, orange, grey and light blue, respectively. $V_g$ and $\Delta$ are the symmetric and antisymmetric combinations of the top and bottom gate voltages. Under a large perpendicular electric field $\propto \Delta$, the contact regions with thick hBN spacer (~10-20 nm) are heavily doped with free electrons and holes, allowing exciton injection and exciton flow through the channel region with thin BN spacer (~2 nm). **c,d,** Schematic for the closed-circuit (**c**) and open-circuit (**d**) drag measurement. A bias current $I_{\text{drive}}$ is applied to the Mo layer; the W contacts are shorted through an ammeter to measure the drag current $I_{\text{drag}}$ in **c**, and in **d** the W contacts are connected to a voltmeter to measure the open circuit drag voltage $V_{\text{drag}}$.

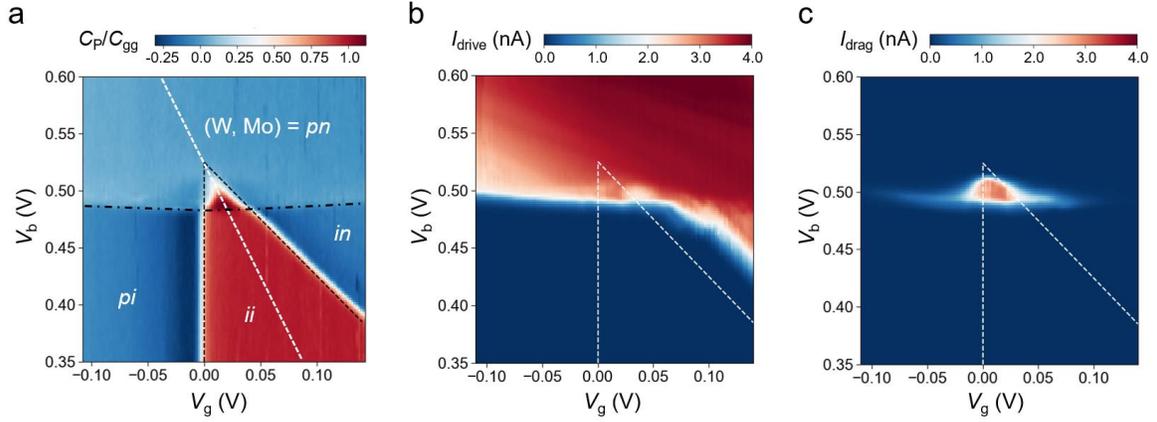

**Figure 2 | Exciton transport and electrostatics phase diagram. a,** Dependence of the normalized penetration capacitance $C_P/C_{gg}$ on $V_b$ and $V_g$, where $C_{gg}$ is the gate-to-gate geometric capacitance. The five distinct regions in the electrostatics phase diagram are separated by the black dashed and dash-dotted lines that mark the TMD band edges (see Ref. [21] for details). Here *i*, *p* and *n* denote, respectively, the intrinsic, hole-doped and electron-doped layer. The system is incompressible in the *ii* region, where both layers are intrinsic, and in the EI region (the triangular area enclosed by the black dashed and dash-dotted lines), where an equilibrium exciton fluid emerges. The white dashed line marks equal electron and hole densities. **b,c,** Drive current in the Mo layer (**b**) and drag current in the W layer (**c**) as a function of $V_b$ and $V_g$. The Mo layer is turned on in the *in*, *pn* and EI regions. Large drag current is observed in the EI region. A weaker drag current is also observed outside the EI region for a slightly electron- and hole-doped EI. All measurements were carried out at T = 1.5 K.

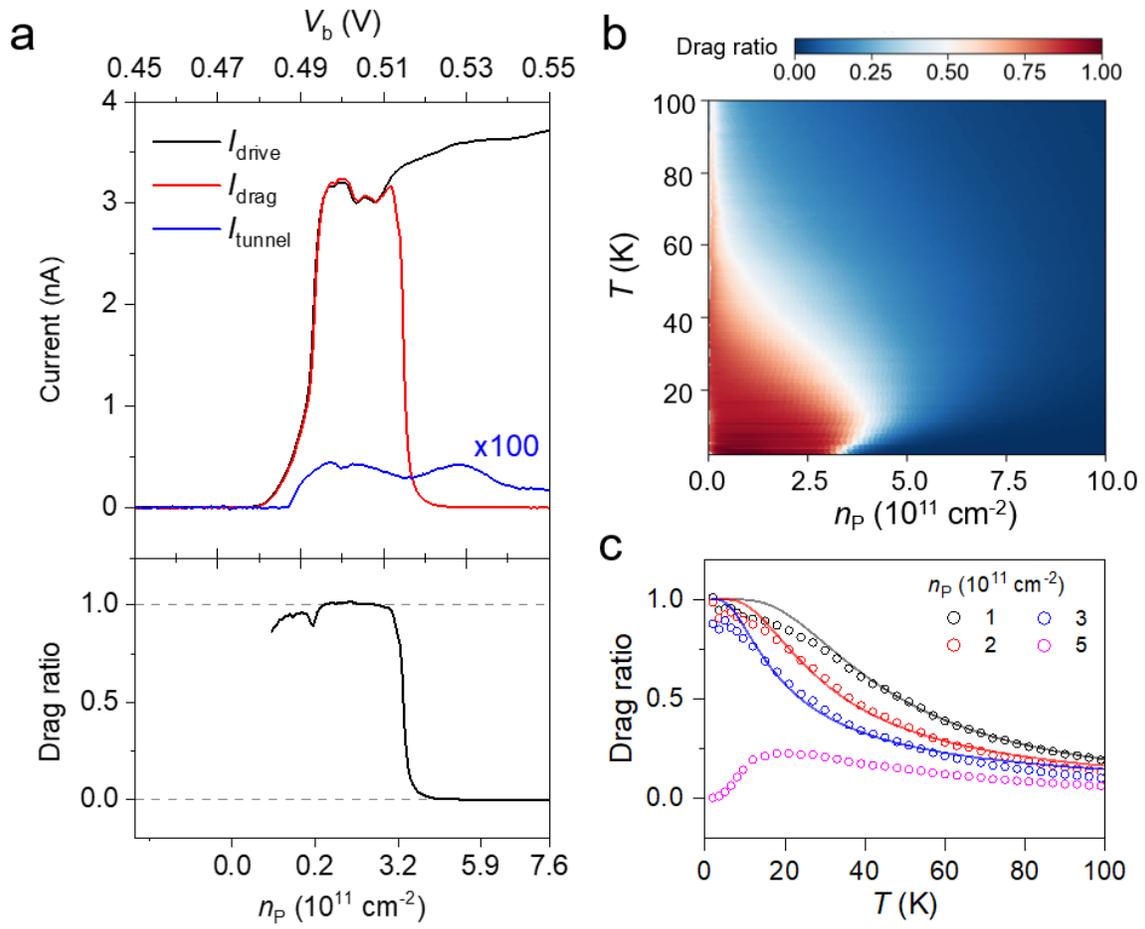

**Figure 3 | Perfect Coulomb drag. a,** Top: Dependence of the drive current, drag current and the tunneling current (multiplied by 100) on $V_b$ at T = 1.5 K. Bottom: The corresponding dependence of the drag current ratio on the pair density. Perfect drag is observed in the EI region, where negligible tunneling current is observed. The drag current ratio quickly drops to zero outside the EI region. **b,** Drag current ratio as a function of the pair density and temperature. **c,** Temperature dependence of the drag current ratio at selected pair densities. Empty dots and solid lines are the experimental data and the theoretical fits described in the main text. Near perfect Coulomb drag is observed at low pair densities and below about 20 K. All data were acquired along the white dashed line in Fig. 2a.

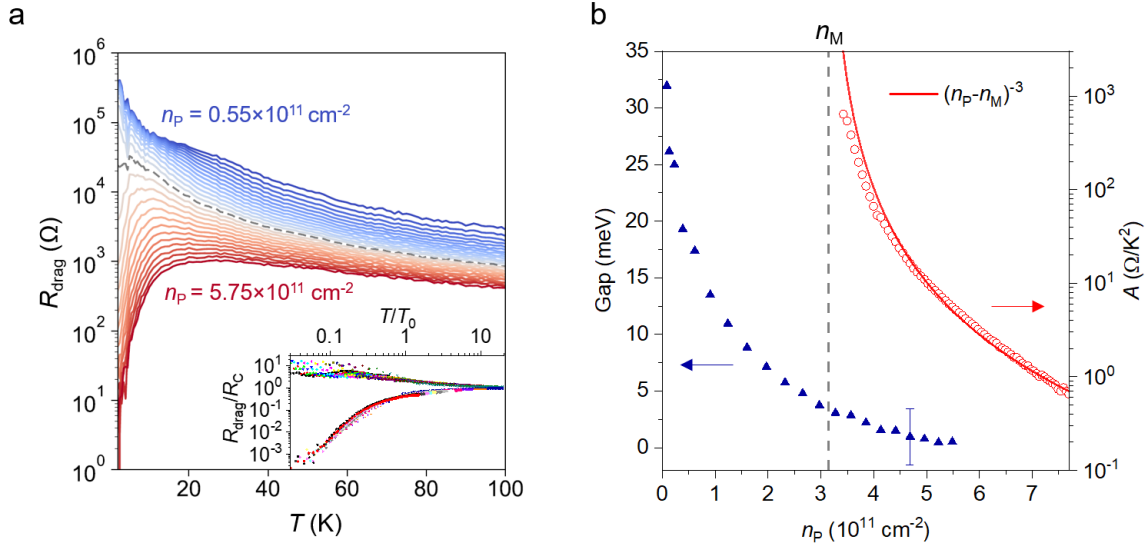

**Figure 4 | Excitonic insulator-to-metal transition. a,** Temperature dependence of the drag resistance at varying pair densities $n_P$. The dashed curve is the critical resistance curve, $R_c$, at the Mott density $n_M \approx 3.2 \times 10^{11}$ cm$^{-2}$, which separates the insulating and metallic behavior at low and high pair densities, respectively. The inset shows the critical scaling of the resistance curves, which collapse into two groups (one metallic and one insulating) around $n_M$. The resistance curves are scaled by $R_c$ at $n_M$ and the temperature is scaled by a density-dependent temperature scale $T_0$. **b,** Dependence of the EI charge gap (solid triangles) and the coefficient $A$ (empty dots) on the pair density $n_P$. The charge gap is obtained from penetration capacitance measurements (see Methods); the error bar represents the systematic uncertainty induced by the finite AC bias voltage in the penetration capacitance measurement. The coefficient $A = R_{\text{drag}}/T^2$ is obtained by fitting the temperature dependent frictional drag at $n_P > n_M$ (Extended Data Fig. 6). The solid red line plots the scaling $A \propto (n_P - n_M)^{-3}$. The vertical dashed line marks the Mott density $n_M$.

**Extended Data Figures**

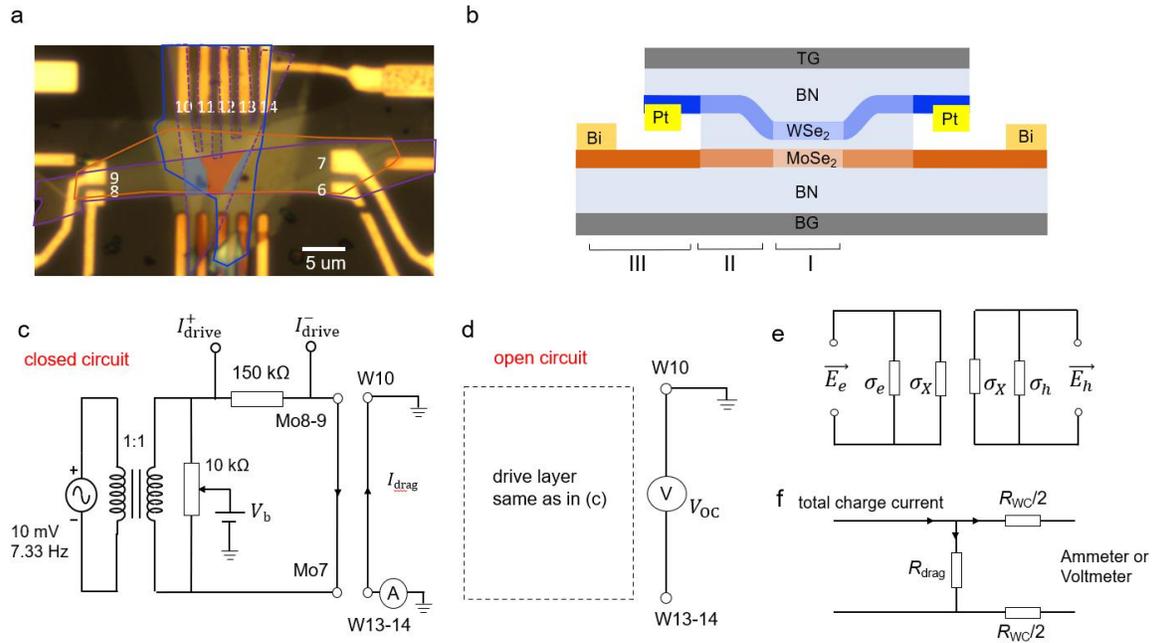

**Extended Data Figure 1 | Coulomb-coupled bilayer device and measurement circuitry. a,** Optical image of the double layer device. The top graphite gate (purple dashed line), bottom graphite gate (purple solid line), W (blue) and Mo (orange) layers are outlined. The blue shaded regions mark the two exciton contacts, and the orange shaded region the excitonic channel. The Pt contacts to the W layer are numbered W10 to W14, and the Bi contacts to the Mo layer are numbered Mo6 to Mo9. **b,** Schematic showing the doping profile in the device. Lighter to darker orange/blue color represents increasing doping densities. Region I is the exciton channel, region II is the exciton contact and region III is the heavily doped metal contact region. The contacts are connected according to the circuit diagrams in **c** (closed-circuit) and **d** (open-circuit). **e,** Simplified circuit model of the bilayer system (left and right block for the electron and hole layer, respectively). There are two parallel conduction channels: $\sigma_X$ for excitons and $\sigma_{e\,(h)}$ for free electrons (holes). **f,** Lumped circuit model in the limit of $\sigma_X \gg \sigma_e, \sigma_D$ (see Methods).

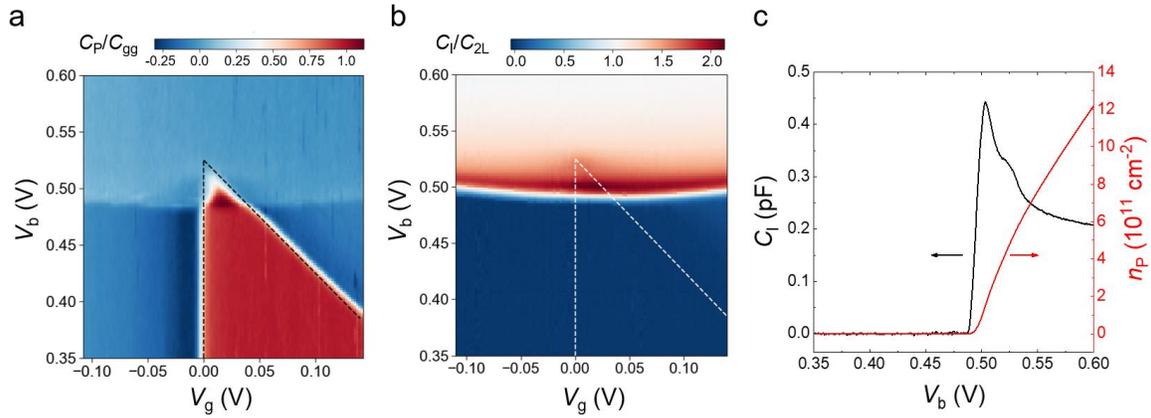

**Extended Data Figure 2 | Capacitance measurements. a,b,** Dependence of the normalized penetration capacitance $C_P/C_{gg}$ (**a**) and interlayer capacitance $C_I/C_{2L}$ (**b**) on $V_b$ and $V_g$ measured at $T = 1.5$K; $C_{2L}$ is the Mo-to-W geometrical capacitance. Both maps are used to determine the electrostatics phase diagram in Fig. 2a. The dashed lines denote the Mo and W band edges. Since $V_b$ is applied to the Mo layer while the W layer remains grounded, the W valance band edge is non-dispersive while the Mo conduction band edge disperses with a slope of -1. **c,** Dependence of $C_I$ (black) and $n_P$ (red) on $V_b$ at equal electron and hole densities. $n_P$ is obtained by integrating $C_I$ with respect to $V_b$.

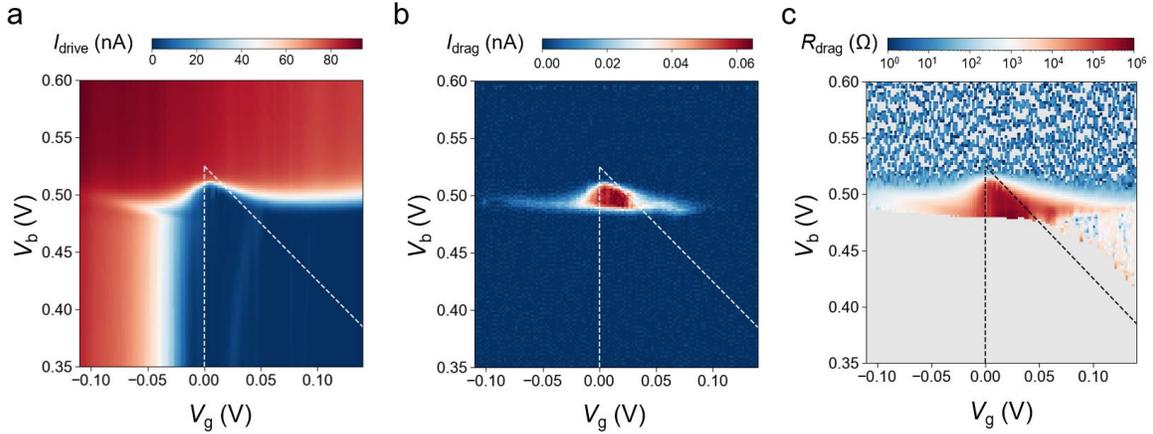

**Extended Data Figure 3 | Additional Coulomb drag measurement results. a,b,** Drive current in the W layer (**a**) and drag current in the Mo layer (**b**) as a function of $V_b$ and $V_g$ at $T = 1.5$ K. The W layer is turned on in the *pi* and *pn* regions. Large drag current is observed in the EI region. Compared to Fig. 2b,c, the large contact resistance in the Mo layer (~ M$\Omega$) results in a substantially suppressed drive current in the EI region. The measurement is therefore similar to an open-circuit measurement (W drive and Mo open circuit) although it is connected in the closed-circuit geometry. Although there is still an exciton current flow in this measurement geometry, perfect Coulomb drag is not observed here because of the effective open-circuit geometry. **c,** Drag resistance as a function of $V_b$ and $V_g$ at $T = 1.5$ K measured by the geometry shown in Fig. 1d. Large drag resistance (~ M$\Omega$) is observed in the EI region. The drag resistance changes by over six orders of magnitude.

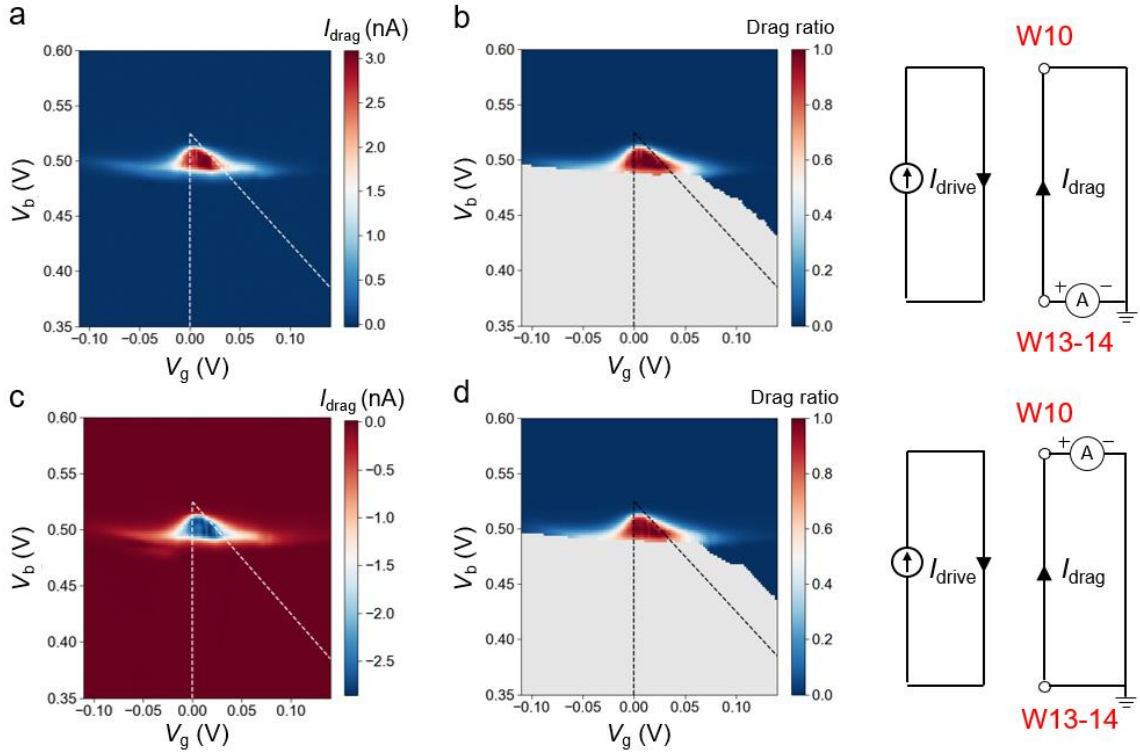

**Extended Data Figure 4 | Determination of the drag current flow direction. a,c,** Drag current in the W layer as a function of $V_b$ and $V_g$ at $T = 1.5$ K measured with the connection shown on the right of the same row. The currents are of opposite sign for the two different measurement geometries. **b,d,** The corresponding absolute drag current ratios are shown. The results here allow us to determine the current flow direction in the W layer and therefore demonstrate that the drive and drag currents flow in the opposite direction in the exciton channel of the device.

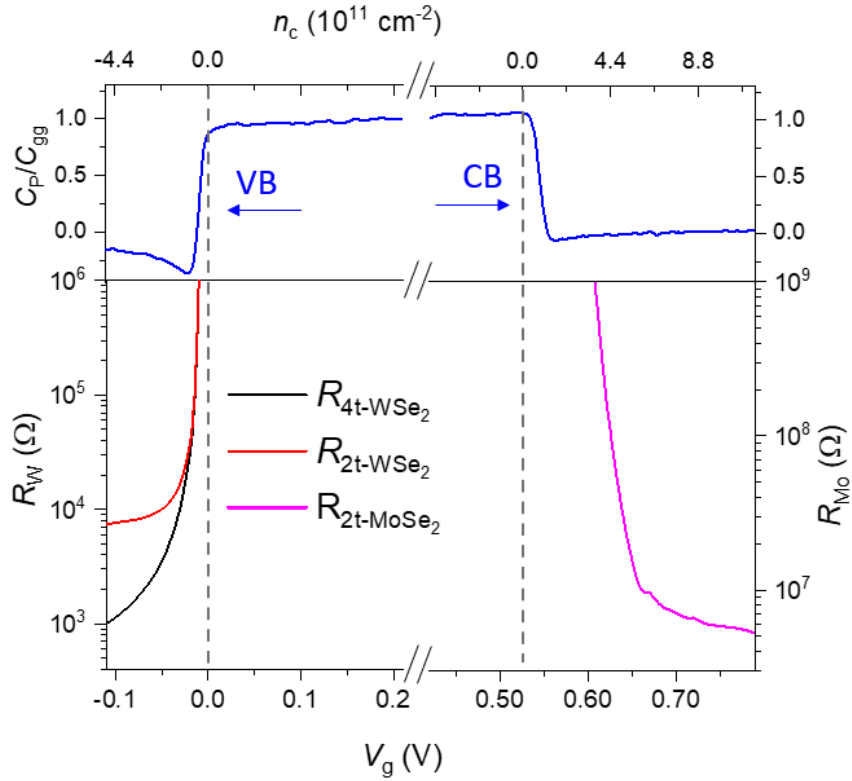

**Extended Data Figure 5 | Transport properties of each TMD layer.** Top: Dependence of the normalized penetration capacitance on $V_g$ (bottom axis) and the free carrier density $n_c$ (top axis). The dashed lines mark the conduction and valence band edges (CB and VB, respectively), beyond which a sharp drop in the penetration capacitance is observed as the system is doped with free electrons/holes. Bottom: The corresponding dependence for the two-terminal (2t) and four-terminal (4t) resistances of each TMD monolayer at zero interlayer bias voltage $V_b = 0$. The results here characterize the transport properties of each TMD layer separately. Only the 2t-resistance can be measured for the Mo layer (pink) while both the 2t- and 4t-resistances can be measured for the W layer (red and black). All of the results here were obtained at $T = 1.5$ K.

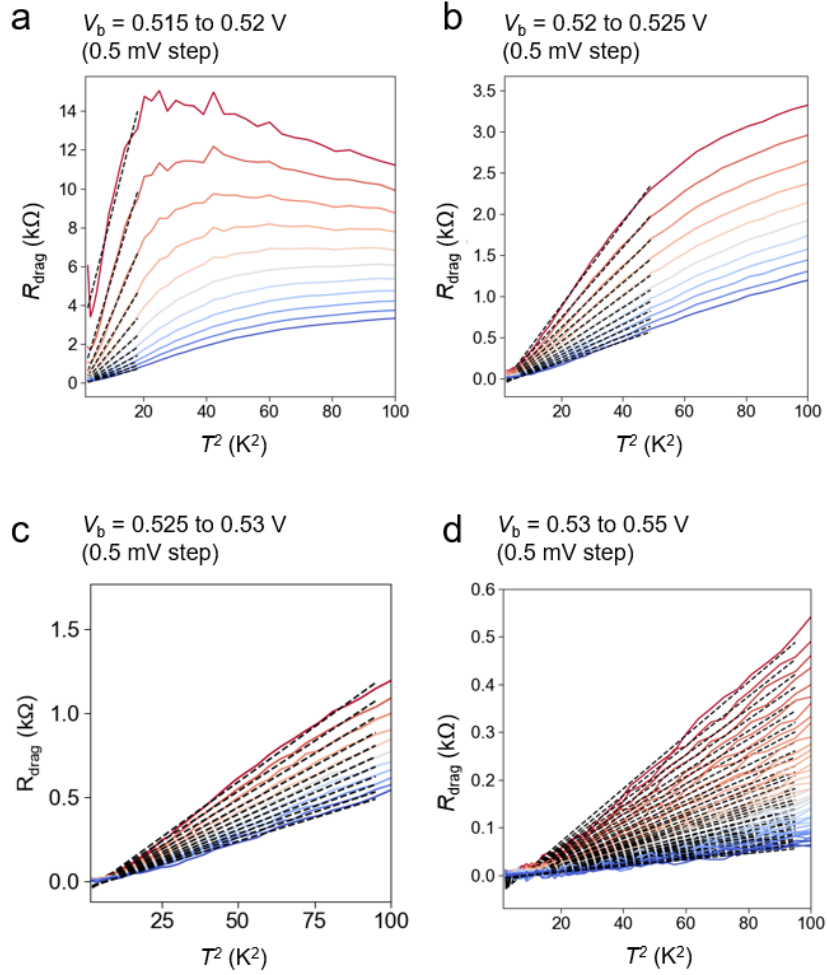

**Extended Data Figure 6 | Frictional drag resistance beyond the Mott density. a-d,** $T^2$ dependence of the Drag resistance at pair densities beyond the Mott density. The different panels show data at different pair density range. For the fitting range $T < 10$ K, the pair density depends only on the bias voltage $V_b$ but not on the temperature. The dashed lines are the linear fits to the data to obtain the coefficient $A$.

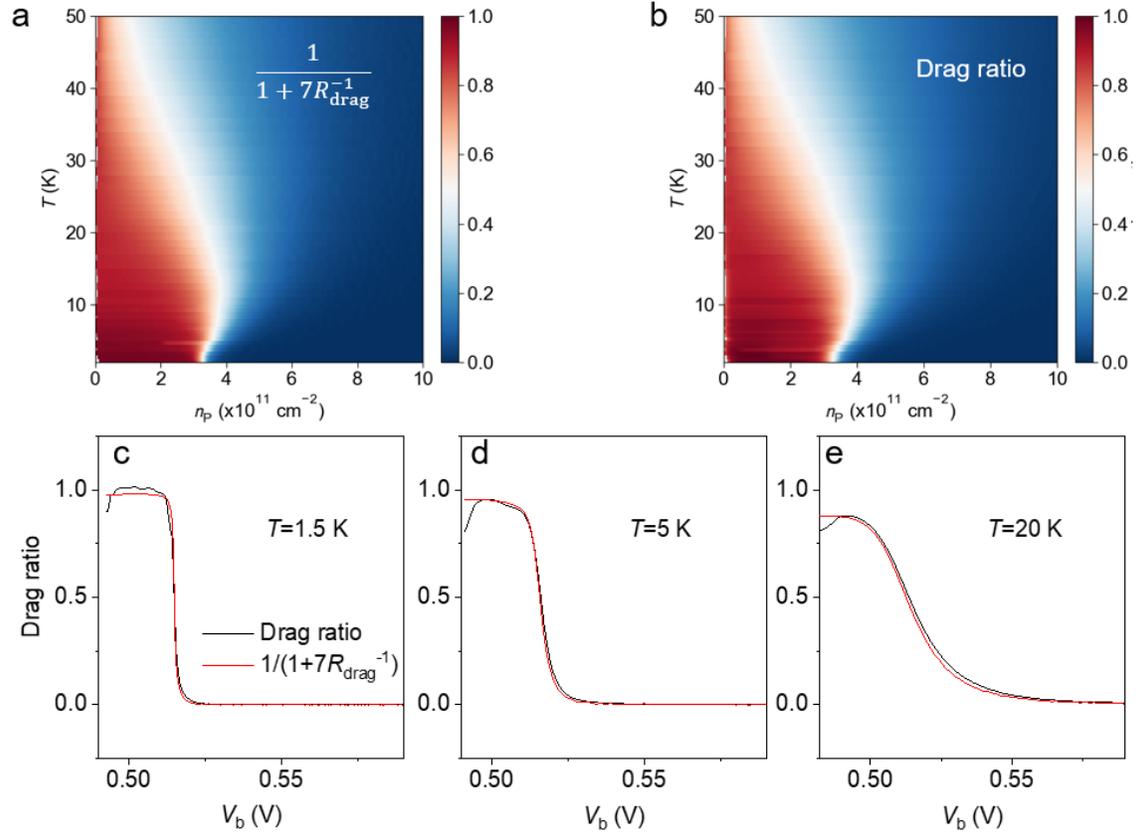

**Extended Data Figure 7 | Relationship between the drag current ratio and the drag resistance. a,** The drag current ratio as a function of the pair density and temperature at equal electron and hole densities calculated from the drag resistance data using the expression $\frac{I_{\text{drag}}}{I_{\text{drive}}} = \frac{1}{1+R_{\text{WC}}/R_{\text{drag}}}$ (with $R_{\text{WC}} = 7$ k$\Omega$). **b,** The actual measured drag current ratio at $T = 1.5$ K (same as Fig. 3b). The calculation in **a** reproduces the actual data in **b**. **c-e,** Dependence of the drag current ratio on the bias voltage at selected temperatures. The black curves are the actual data and the red curves are the results reproduced from the calculation $\frac{I_{\text{drag}}}{I_{\text{drive}}} = \frac{1}{1+R_{\text{WC}}/R_{\text{drag}}}$.

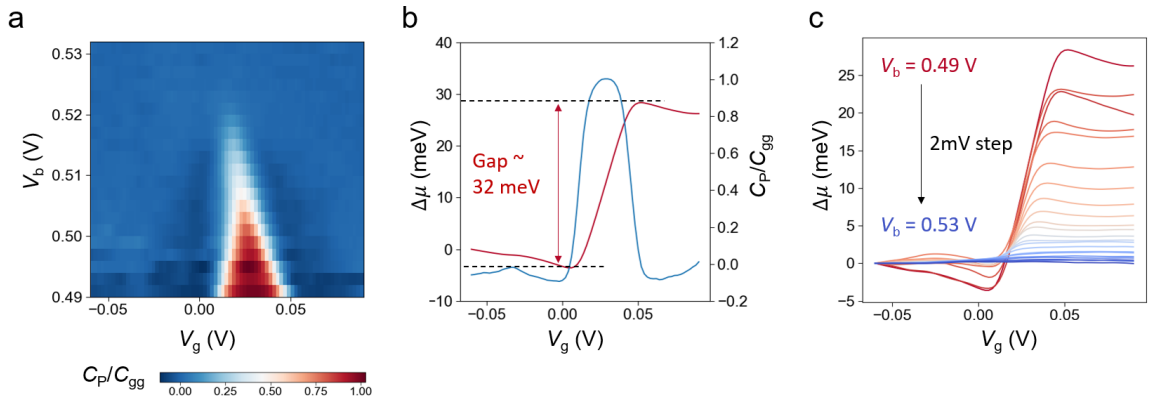

**Extended Data Figure 8 | Determination of the charge gap. a,** Zoom-in view of the normalized penetration capacitance as a function of $V_b$ and $V_g$ at 1.5 K. The red triangle is the EI region. **b,** Dependence of the normalized penetration capacitance (blue) and the calculated charge chemical potential $\Delta\mu \approx e\int \left(\frac{C_P}{C_{gg}}\right) dV_g$ (red) on $V_g$ at $V_b = 0.49$ V. A charge gap of about 32 meV can be obtained by the size of the chemical potential jump (bound by the dashed lines). **c,** Dependence of $\Delta\mu$ on $V_g$ at varying $V_b$ from 0.49 V to 0.53 V. A continuously vanishing charge gap with increasing $V_b$ is observed.

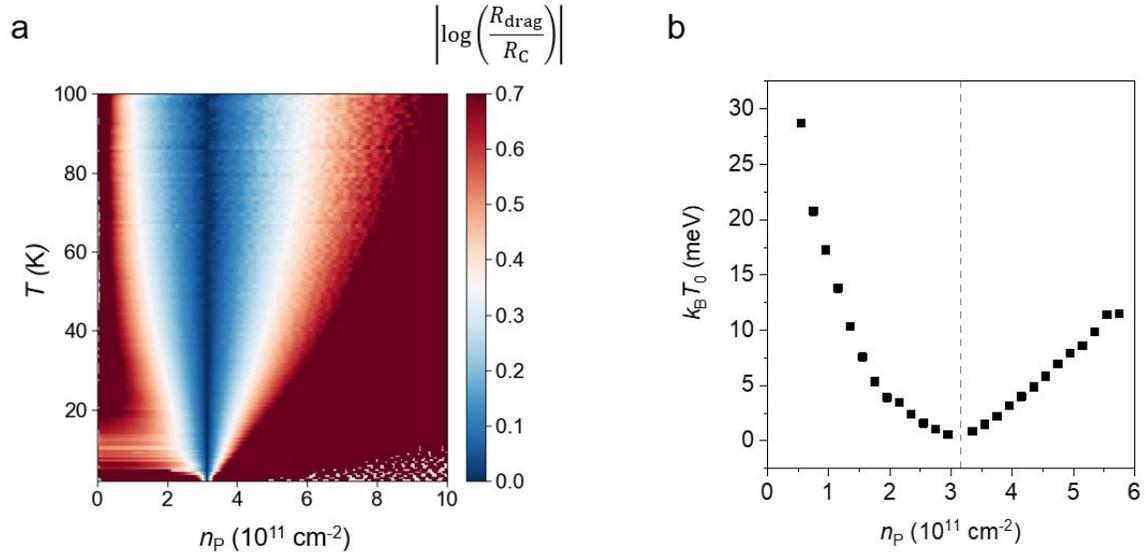

**Extended Data Figure 9 | Critical scaling analysis of the excitonic insulator-to-metal transition. a,** Scaled drag resistance as a function of the pair density and temperature in log scale. It suggests a quantum critical fan around the Mott density $n_M = 3.2 \times 10^{11}$ cm$^{-2}$. **b,** Pair density dependence of the temperature scale $T_0$ that collapses the resistance curves in the inset of Fig. 4a. It continuously vanishes towards the Mott density (dashed line) from both sides.